\documentstyle[prl, multicol, aps, epsf]{revtex}

\begin{document}
\draft

\title
{\large \bf Quantum Tunneling and Quantum-Classical Transitions 
in Large Spin Systems}

\author{
Bambi Hu$^{1, 2}$, De-gang Zhang$^{1, 3}$, and Bo-zang Li$^{4}$}

\address
{$^{1}$Department of Physics and Centre for Nonlinear Studies,
Hong Kong Baptist University, \\Kowloon Tong, Hong Kong, China\\
$^{2}$Department of Physics, University of Houston, Houston, 
TX 77204, USA\\
$^{3}$Institute of Solid State Physics, Sichuan Normal University, 
Chengdu 610068, China\\
$^{4}$Institute of Physics, Chinese Academy of Sciences, 
P. O. Box 603, Beijing 100080, China}

\maketitle

\begin{abstract}

We have studied quantum tunneling of large spins in a biaxial 
spin system and in a single-axial spin system with a transverse 
magnetic field. The asymptotically exact eigenvalues and 
eigenstates of the spin systems are obtained by solving the Mathieu 
equation. When the height $4q$ of energy barrier formed by the 
anisotropy and the magnetic field exceeds some 
critical values $4q_{1T}(r)$, the ground states and the excited 
states split due to tunneling of large spins. We also have 
presented the phase diagram of these spin systems in the extremely 
weak spin-phonon coupling limit. The crossover from thermal to 
quantum regime is second order. With further decreasing temperature,   
the first order phase transition occurs from a thermally-assisted 
resonant tunneling to thermal regime as $q<q_{1T}(s)$. Our theory 
agrees with recent experimental observations well.

\end{abstract}

\begin{multicols}{2}

In recent years quantum tunneling of magnetization (QTM) in small
magnetic particles has been widely investigated in both theory and
experiment because of its fundamental interest in exploring the 
transition between classical and quantum
 physics [1]. It gives 
evidence of quantum mechanical behavior in macroscopic systems. 
Such phenomena have been observed experimentally in ferritin [2], 
the molecular nanomagnets Mn$_{12}$-ac [1, 3, 4], Fe$_8$ [5, 6], 
BaFeCoTiO [7], etc at very low temperatures. Theoretical calculations  
of such effects were performed for the ferromagnets [1, 8] and 
antiferromagnets without and with the excess spin [1, 9] 
by using path integral and WKB methods, etc.. However, these approaches 
depend on the semiclassical treatment and did not present the energy spectrum 
of the spin systems, which is necessary to give the complete solutions
of spin tunneling problems. Especially, the quantization of spin levels 
becomes very important to explain well the experiments of quantum tunneling 
at very low temperatures [4]. In very recent Refs. [10-12], the spin 
systems arouse new interest again because they provide examples exhibitting 
first order phase transitions, which were not known previously.  
The energy spectrum of the spin systems can also help us to understand the
origin of the first order phase transition.
In this paper, we investigate a biaxial spin system without the magnetic
field and a single-axial spin system
 with a transverse magnetic field 
which are generic for spin tunneling problems studied by different methods 
[1, 8, 10-12] in the framework of the Schr$\stackrel{..}{o}$dinger's 
picture of quantum mechanics, so that the energy spectrum as well as the 
physical properties of the spin systems are obtained.

We first consider QTM in the biaxial spin system without the magnetic
field described by the Hamiltonian [1, 8, 11]
$$H=AS^2_z-BS^2_x,\eqno{(1)}$$
where the anisotropy constants $A$ and $B$ are taken to be positive.
So the ground state of the system corresponds to spin ${\bf S}$ 
pointing the positive or negative x axis. 
To diagonalize the Hamiltonian
(1), we first write its matrix representation in the basis $|s, m>$, which
has the properties: ${\bf S}^2|s, m>=s(s+1)|s, m>, S_z|s, m>=m|s, m>$ 
and $S_{\pm}|s, m>\equiv (S_x\pm iS_y)|s, m>=
\sqrt{s(s+1)-m(m\pm 1)}|s, m\pm 1>$, 
where $s$ is the spin quantum number to be taken as an integer [13], 
$m=-s, -s+1, \cdots, s$ and the unit of $\hbar=1$ is used.
Let $E$ and $\Psi_m$ be the eigenenergies and the eigenstates of $H$, 
respectively, then we have
$$\begin{array}{c}
\sqrt{[s(s+1)-(m-1)^2]^2-(m-1)^2}\Psi_{m-2}\\
+\sqrt{[s(s+1)-(m+1)^2]^2-(m+1)^2}\Psi_{m+2}\\
+4[\frac{E}{B}+\frac{1}{2}s(s+1)
-(\lambda+\frac{1}{2})m^2]\Psi_m=0,
\end{array}\eqno{(2)}$$
where $\lambda=\frac{A}{B}$.
For a large spin system, $s$ is a sufficiently large number, i.e. $s\gg1$. 
Define $\Phi_m=(-1)^{[\frac{m}{2}]}\Psi_m$, where $[\frac{m}{2}]$ denotes 
the integer part of $\frac{m}{2}$,then Eq. (2) becomes
$$(1-x^2)\frac{d^2\Phi}{dx^2}-2x\frac{d\Phi}{dx}+[-\frac{E}{B}
-\frac{1}{4}+\lambda s(s+1)x^2-\frac{1}{4}\frac{1}{1-x^2}]\Phi 
=0 \eqno{(3)}$$
in the large-$s$ limit. Here $x=\frac{m}{\sqrt{s(s+1)}}$ and only
the leading terms are remained. Taking the transformations $\Phi=
(1-x^2)^{-\frac{1}{4}}y(x)$ and $x=\sin t$ and substituting them into
Eq. (3), we finally obtain the well-known Mathieu equation 
$$\frac{d^2y}{dt^2}+[\Lambda(q)-2q\cos (2t)]y=0,\eqno{(4)}$$
which describes the motion of a particle with the mass $\frac{1}{2}$
in the cosine potential. Here the characteristic values $\Lambda(q)=
-\frac{E}{B}+2q$ and $q=\frac{1}{4}\lambda s(s+1)$.
Obviously, when $q=0$, $\Lambda_{|m|}(0)=m^2$, then $E_m=-Bm^2$,  
which are the exact eigenenergies of $(1)$ in the eigenstates $|s, m>_x$.
Now we have completed the mapping of the spin problem onto a particle
problem. The energy spectrum of the Hamiltonian (1) is determined 
completely by the Mathieu equation.

The Mathieu equation (4) has been studied in a large set of literature
[14, 15] due to its physically basic importance. It is known that there 
exist periodic solutions of period $n\pi$, where $n$ is any positive integer. 
However, the solutions relative to quantum tunneling problem discussed here  
are only those even and odd ones with periods $\pi$ and $2\pi$, 
i.e., $y=$ce$_{r}(t, q)$ and se$_r(t, q)$, $r=0, 1, \cdots, s$.
Assume that $a_r$ and $b_r$ 
are the characteristic values associated with the even and odd periodic solutions, 
respectively, then these characteristic values  
form a countable sequence, i.e. 
$a_0<b_1<a_1<b_2<a_2<\cdots<b_s<a_s$ for $q>0$ and 
$a_{2i}(-q)=a_{2i}(q)$, $b_{2i}(-q)=b_{2i}(q)$, $a_{2i+1}(-q)=b_{2i+1}(q)$ 
and $b_{2i+1}(-q)=a_{2i+1}(q)$. 
We note that the approximately degenerate characteristic levels 
$(a_r, b_r)$ are lifted remarkably as the energy barrier parameter $q$ exceeds 
the critical value $q_{1T}(r)$. The higher the characteristic level is, the
larger the critical value is, i.e. $q_{1T}(i)>q_{1T}(j)$ for $i>j$.
The value of $q_{1T}(r)$ can be estimated by letting the height 
of the energy barrier be equal to the energy $\Lambda_{r}(0)$ of the particle 
in the vanishing energy barrier, i.e. 
$4q_{1T}(r)=r^2$, which is a good approximation for large $r$. This shows
clearly that when $q>q_{1T}(r)$, the splitting of the characteristic levels
$(a_r, b_r)$ is induced by {\it tunneling} of the particle.
As $q$ is larger than $q_{2T}(r)\approx \frac{[s(s+1)]^2}{4(s-r+1)^2}$ [16], 
the levels $b_r$ and 
$a_{r-1}$ degenerate approximately due to very high energy barrier.  
Therefore, in order to observe experimentally tunneling splitting
of level $r$, $q$ must locate between $q_{1T}(r)$ and $q_{2T}(r)$.
We also note that tunneling of the particle mainly occurs in those 
characteristic levels ranging from the highest level to one with 
characteristic value $a_{s}-s^2+0(s)$ as $q<q_{1T}(s)$, which is important 
to evaluating approximately the critical temperature from thermal to 
quantum regime below.    
                                                   
For the Hamiltonian (1), its lower excited  states correspond to the higher
characteristic levels of the Mathieu equation. In order to see the tunneling 
splitting of the degenerate ground states, 
$q$ must exceed the critical value $q_{1T}(s)\approx \frac{1}{4}s^2$, i.e. 
$\lambda_{1T}(s)\approx 1-\frac{1}{s}$, which coincides with the result 
reported previously $[11]$ when $s$ is very large. When $q>q_{2T}(m)$,
we obtain tunneling splitting of the ground states
as well as the excited states $[14]$
$$\Delta E_m=B(a_m-b_m)=B(2\sqrt{\lambda s(s+1)}-m)+
0(q^{-\frac{1}{2}}).\eqno{(5)}$$
When $q\rightarrow\infty$, the ground state $a_s$ is
singlet while all the excited states are twofold degenerate, which
coincide with the degeneracy of the Hamiltonian (1) with $B\rightarrow 0$.

Our another example is devoted to the single-axial spin system
with a transverse magnetic field, which Hamiltonian has form [1, 8, 10]
$$H=-hS_z-BS_x^2,\eqno{(6)}$$
where $B>0$ and $h>0$. The Hamiltonian (6) can be also diagonalized by
the same method mentioned above. Its energy spectrum is determined by the 
Mathieu quation (4), too. However, in this case, $x=\cos(2t)$, 
$\Lambda(q)=-\frac{4E}{B}$ and $q=\frac{2h\sqrt{s(s+1)}}{B}$. 
As $h=0$, then $\Lambda_r(0)=r^2$ and $E_m=-Bm^2$. So the ground states and 
the excited states of the Hamiltonian (6)
correspond to the characteristic levels $(a_{2s}, b_{2s})$ and
$(a_{2r}, b_{2r}) (r<s)$, respectively. According to that 
$4q_{1T}(m)=(2m)^2$, we have the critical magnetic field 
$h_{1T}(m)=\frac{Bm^2}{2\sqrt{s(s+1)}}, m\gg 1$. As $q<h_{1T}(m)$, 
there is no tunneling between the approximately degenerate energy levels 
$E_m$ (i. e. $a_{2m}$ and $b_{2m}$) of the Hamiltonian (6).
As $h>h_{1T}(s)=\frac{Bs^2}{2\sqrt{s(s+1)}}$, the ground states occur 
splitting, which is the same with that
in Ref. [10] for very large $s$. We also note that $h$ must be smaller than 
$h_c=2Bs$, at which the degenerate ground states coincide.
As $q\gg 1$, we have
 
$$\Delta E_m=\frac{1}{4}B(a_{2m}-b_{2m})=B\Big(\sqrt{\frac{2h
\
\sqrt{s(s+1)}}{B}}-\frac{1}{2}m\Big)+0(q^{-\frac{1}{2}}).
\eqno{(7)}$$
When $q\rightarrow\infty$, all the energy levels $a_{2m}$ and $b_{2m}$
are singlet, which agree with the energy spectrum of the Hamiltonian (6) 
with $B\rightarrow 0$. Because there is no Kramers' degeneracy in the
Hamiltonian (6), tunneling splitting also holds for half odd integer 
spin $s$ [14].

Up to now, we have constructed the energy spectrum of the pure spin systems 
(1) and (6). However, the experiments of observing quantum tunneling 
were performed at low temperatures. So the influence of phonons or
thermal activation on 
tunneling must be considered. Here, we assume the spin-phonon coupling
is so weak that the energy spectrum of the spin systems is not changed too much.
The transition between the ground state and the excited states can be 
completed by absorbing or emitting phonons. With the help of the Debye theory
of a solid [17], we obtain approximately the critical temperature $T_c$ 
satifying 
$$\begin{array}{rcl}
3k_BT_cD(\frac{T_c}{\theta})&\approx&Bs^2
\end{array}\eqno{(8)}$$
as $q<q_{1T}(s)$ and 
$$\begin{array}{rcll}
3k_BT_cD(\frac{T_c}{\theta})&\approx&B(a_s-b_r)&{\hbox {for (1)}}\\
&\approx&\frac{1}{4}B(a_{2s}-b_{2r})&{\hbox {for (6)}}
\end{array}\eqno{(9)}$$
as $q_{2T}(r)\leq q\leq q_{2T}(r+1)$ and $q_{2T}(2r)\leq q\leq q_{2T}(2r+1)$,
respectively. 
Here the Debye function $D(x)\approx \frac{1}{5}\pi^4x^3$ at low temperature, 
$k_B$ and $\theta$ are the Boltzmann constant and the Debye temperature, 
respectively. Obviously, the crossover 
from the high temperature (thermal) phase to the low temperature (quantum) 
phase is second order because tunneling splitting of large spin  
continuously transforms into zero. As $q<q_{1T}(s)$,
in order to see tunneling, large spin must absorb at least energy $B(s^2-4q)$
and $B(s^2-q)$ for the models (1) and (6), respectively, provided mainly by
phonons or thermal activation.
So we have the first order phase transition temperature $T_0$, 
$$\begin{array}{rcll}
3k_BT_0D(\frac{T_0}{\theta})&\approx&B(s^2-4q)&{\hbox {for (1).}}\\
&\approx&B(s^2-q)&{\hbox {for (6).}} 
\end{array}\eqno{(10)}$$
The phase diagram of the spin systems $(1)$ and $(6)$ is shown in Fig. $1$.

Here we compare our theoretical results with recent experiments of quantum
tunneling. The Hamiltonian $(1)$ is believed to be  a good description
for the molecular nanomagnet Fe$_8$ with spin $s=10$ [5]. The parameters
$A=0.092$K and $B=0.224$K, then $q=11.295<q_{1T}(s)=25$. Therefore,
with decreasing temperature, the spin system enters the 
thermally-assisted resonant tunneling regime II from thermal regime I, as observed
down to $0.067$K. However, tunneling can not be seen possibly if the temperature
is further lowered to below $T_0$, which would be of interest to be 
verfied experimentally. The Hamiltonian (6) has been found to be a good
approximation for the molecular nanomagnet Mn$_{12}$-ac with spin 
$s=10$ [4]. For this nanomagnet, $B=0.68$K, then $h_{1T}(s)=3.24$K. The parameter
$h=g_{\perp}\mu_{B}H$ was changed from 6.38K to 8.93K in experiment, i.e.
the magnetic field $H$ was applied between 5T and 7T. So the spin system fell 
into quantum regime III, where the two states of the fundamental doublet
of the high-spin molecule are lifted. The transitions between the two
states induced by a two-phonon process have been observed from 0.1K to
0.02K when $H=6.06$T [4]. Obviously, as the spin systems just set in the
thermal regime IV or $q>q_{2T}(s)$, tunneling does not be observed in 
experiments even at very low temperatures [18].

In summary, we have investigated quantum tunneling of large spins
based on simple models (1) and (6). The asymptotically exact spectrum
of the spin systems is completely determined in the full range of the 
magnetic anisotropy and the magnetic field. The degeneracy of all
the energy levels is removed due to QTM. The phase diagram presented 
here coincides with the experimental observations. 
To our knowledge, the thermal regime IV in Fig. 2 was not predicted by
previous theories. We think that
the first order phase transition is not limited to the spin systems
(1) and (6) and also exists generally in other spin systems, depending on
whether the energy barrier and the ground states as well as the
low-lying excited states cross. The method of
solving the spin Hamiltonians used above is also applied to the other
large spin systems, such as the other symmetric ferromagnets, 
antiferromagnets and etc., which is in progress.        

This work is supported in part by grants from the Hong Kong Research
Grants Council (RGC), the Hong Kong Baptist University Faculty
Research Grant (FRG), the Sichuan Youth Science and Technology 
Foundation, the NSF of the Sichuan Educational Commission
and the NSF of China (No. 19677101).

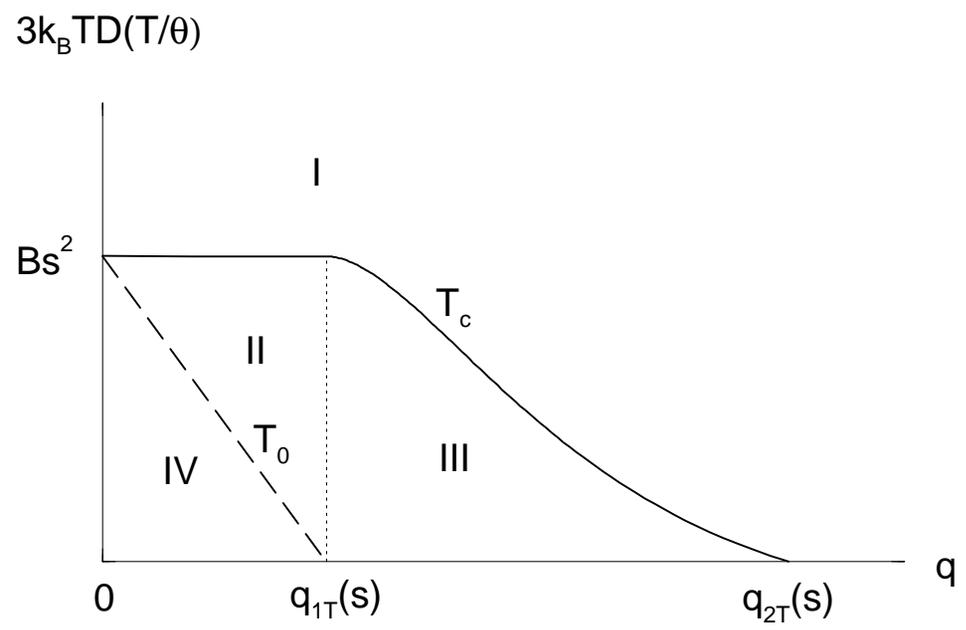
\begin{figure}
\epsfxsize=8cm
\vspace{2cm}
\narrowtext
\caption{The phase diagram of the spin systems (1) and (6): I and IV are the
thermal regimes; II is the thermally-assisted resonant tunneling regime; 
III is the quantum tunneling regime. The transition from I to II and III is 
second order while the transition from II to IV is first order.}
\end{figure} 

\end{multicols}


\begin{thebibliography}{99}
\bibitem{16} The values of $q_{2T}(r)$ can be obtained by diagonalizing the
Hamiltonian (1) in the basis $|s, m>_x$, which also leads the Mathieu equation.
$q_{1T}(r)$ and $q_{2T}(r)$ coexist.
\end{thebibliography}
\end{document}